\documentclass[useAMS,usenatbib]{mn2e}

\usepackage{graphicx}
\usepackage{txfonts}
\usepackage[normalem]{ulem}
\usepackage[usenames,dvipsnames]{color}

\newcommand{\fig}[1]{Fig.~\ref{#1}}

\def\spose#1{\hbox to 0pt{#1\hss}}
\def\gsim{\mathrel{\spose{\lower 3pt\hbox{$\mathchar"218$}}
          \raise 2.0pt\hbox{$\mathchar"13E$}}}
\def\lsim{\mathrel{\spose{\lower 3pt\hbox{$\mathchar"218$}}
          \raise 2.0pt\hbox{$\mathchar"13C$}}}

\newcommand{\lbol}{L_{\rm bol}}
\newcommand{\erad}{\epsilon_{\rm e}}
\newcommand{\ee}{\epsilon_{\rm e}}

\def\tf{t_{\rm f}}
\def\dtf{\Delta t_{\rm f}}
\def\betars{\beta_{\rm RS}}

\def\Gr{\Gamma_{\rm rel}}
\def\Gej{\Gamma_{\rm ej}}
\def\Gbw{\Gamma_{\rm bw}}
 \def\pIFS{p_{\rm IFS}}
 \def\pRS{p_{\rm RS}}
 \def\msh{m_{\rm sh}}
 \def\tcross{t_{\rm cross}}
 \def\tdiss{t_{\rm diss}}

\title[X-ray flares from dense shells]
{X-ray flares from dense shells 
formed in gamma-ray burst explosions}
\author[R. Hasco\"et et al.]{R. Hasco\"et$^{1}$\thanks{E-mail:
hascoet@astro.columbia.edu}, A. M. Beloborodov$^{1}$, F. Daigne$^{2}$, R. Mochkovitch$^{2}$\\
$^{1}$Physics Department and Columbia Astrophysics Laboratory, Columbia University, 538 West 120th Street, New York, NY 10027, USA.\\
$^{2}$UPMC-CNRS, UMR7095, Institut d'Astrophysique de Paris, F-75014, Paris, France.}
\begin{document}

\date{Accepted **.**.**. Received **.**.**; in original form **.**.**}

\pagerange{\pageref{firstpage}--\pageref{lastpage}} \pubyear{2014}

\maketitle

\label{firstpage}

\begin{abstract}
Bright X-ray flares are routinely detected by the {\it Swift} satellite 
during the early afterglow of gamma-ray bursts, when the explosion ejecta 
drives a blast wave into the external medium. We suggest that the flares 
are produced as the reverse shock propagates 
into the tail of the ejecta. The ejecta is expected to contain
a few dense shells formed at an earlier stage of the explosion.
We show an example of how such dense shells form and describe how the reverse shock 
interacts with them. A new reflected shock is generated in this interaction,
which produces a short-lived X-ray flare.
The model provides a natural explanation for the main observed features of the 
X-ray flares --- the fast rise, the steep power-law decline, and the characteristic peak duration 
$\Delta t /t= (0.1-0.3)$. 
\end{abstract}


\maketitle

\section{Introduction}
\label{sect_intro}

About 30\% of GRBs show X-ray flares during their early afterglow
(e.g. \citealt{burrows_2005b,falcone_2007,chincarini_2010}).
Sometimes they are accompanied by significant flux increase in the optical band
(e.g. \citealt{li_2012}).
The X-ray flares are characterized by a fast rise 
of luminosity (typically by a factor $\sim 10$) 
to a sharp peak followed by a 
 power-law decay \citep{chincarini_2007}.
The ratio of the characteristic temporal width of the peak $\Delta t_{\rm f}$ 
to its time of occurrence since the beginning of the GRB
is typically $\Delta t_{\rm f} / t_{\rm f} \sim 0.1-0.3$.
At the end of the flare, the X-ray flux resumes the underlying smooth decay of the early afterglow,
with no apparent flux increment left.

Besides the X-ray flares, the early afterglow shows other puzzling features, such 
as plateaus and sudden steep decrease in the observed luminosity.
These features are not explained by the standard forward shock model 
 \citep{meszaros_1997, sari_1998}, and it was proposed that the observed afterglow
 is produced by a long-lived reverse shock inside the GRB ejecta
\citep{uhm_2007, genet_2007}. 
The dynamics and emission of the reverse shock are particularly sensitive to 
the structure of the ejecta (the distribution of its Lorentz factor, density, and magnetic fields)
which may explain the rich
phenomenology of the early afterglow
 \citep{hascoet_2011, hascoet_2012, uhm_2012, hascoet_2014}.

The fast rise and relatively short duration of the X-ray flares motivated several authors
to invoke late activity of the central engine and associate the flares with
the jet produced by this activity --- a scaled-down version of the prompt
GRB emission (e.g. \citealt{burrows_2005, fan_2005, zhang_2006}).
This possibility requires a mechanism that keeps the central engine active for about one day,
as some flares are observed as late as a day after the GRB explosion. It also requires 
the engine to be variable in a special way on a timescale 
$\Delta t_{\rm f} \sim (0.1-0.3) t_{\rm f}$, in contrast to 
the much faster, multi-peak variability observed during the prompt GRB.

This Letter proposes an alternative possibility
that the flares are produced by the long-lived reverse shock 
when it crosses the tail of the GRB ejecta.
Our model does not 
require a long-lived central engine; it requires that the Lorentz factor in the end 
of the GRB explosion is significantly reduced, from 
$\Gamma\sim 100$ to $\Gamma\sim 10$, which leads to the formation
of an extended tail (e.g. \citealt{uhm_2007, genet_2007}). 
As described in Section~\ref{sect_ejecta_structure}, dense shells are expected to form in the tail of the 
expanding ejecta before the passage of the reverse shock. In Section~\ref{sect_dissPow}
we estimate the bolometric light-curve 
of emission produced by the encounter of the reverse shock with the dense shell,
and find that it resembles 
the observed X-ray flares. Comparison of the model with observations and future 
possible ways of its development are discussed in Section~\ref{sect_discussion}.

\section{Dense shells after internal shocks}
\label{sect_ejecta_structure}

The observed variability of GRB emission
\citep{fishman:95,beloborodov_2000, guidorzi_2012}
suggests that the Lorentz factor of the relativistic ejecta $\Gamma$
fluctuates in a broad range of timescales, from a few milliseconds to several minutes.
In the expanding outflow, regions where the radial gradient of 
Lorentz factor is negative (i.e. where $\Gamma$ is decreasing outwards) 
are progressively compressed. 
If the ejecta is not magnetically dominated, internal shocks eventually form and propagate.
The internal shocks can impact the prompt $\gamma$-ray emission mainly in two ways:
\textit{(i)} in the subphotospheric (optically thick) region, shock heating can offset the 
adiabatic cooling of radiation and change its spectrum from thermal to 
the observed Band-type shape through the Comptonization process and 
additional synchrotron emission
\citep{meszaros_2000, peer_2006, beloborodov_2010, vurm_2011};
\textit{(ii)} 
outside the photosphere, shocks can continue to produce synchrotron emission 
in the gamma-ray band if they efficiently accelerate particles \citep{rees_1994, kobayashi_1997, daigne_1998}.

The outflow is likely initially inhomogeneous, and 
internal shocks can amplify the density variations, especially if the shocks are radiatively 
efficient. More importantly, the outflow tends 
to develop a few massive dense shells long {\em after} the shocks. At time $t_0$
when the internal shock phase ends, the 
fluctuations of Lorentz factor $\Gamma$ 
have been damped and $\Gamma$ acquires a monotonic radial profile. 
Its slope $d\Gamma/dr>0$ is significantly inhomogeneous, 
as illustrated below by a hydrodynamical simulation.
The subsequent ballistic expansion of the outflow at $t\gg t_0$ 
generates a large contrast in density, as can be seen from the following estimate.

Consider a shell of width $\Delta_{\rm sh}$ with the  
Lorentz factor variation across the shell 
$\delta\Gamma=(d\Gamma/dr)\,\Delta_{\rm sh}$.
As the shell expands ballistically, its width evolves as
\begin{equation}
  \frac{\Delta_{\rm sh}(t)}{\Delta_0} = 1 + \frac{\delta \Gamma}{\Gamma} \frac{t}{t_0},
\end{equation}
where $\Delta_0$ is the shell width at $t_0$.
The relevant $\Delta_0$ at the end of the internal shock phase corresponds to 
the longest variability timescale observed in the burst, $\Delta_0\sim ct_0$, which 
is comparable to the burst duration, e.g. $t_0 \sim 1-10$~s.
In contrast, the time $t$ given to ballistic expansion before the reverse 
shock crosses the outflow in our model is associated with the observed time of the X-ray flares,
$t  \sim t_f \sim 10^3-10^4$~s.

The large ratio $t/t_0$ implies that $\delta\Gamma$ should strongly 
affect the density structure of the outflow. The parts of the outflow with 
$\delta\Gamma\sim\Gamma$ will become much thicker than the parts with 
$\delta\Gamma\ll\Gamma$, which will form dense shells. The corresponding density 
contrast is roughly given by
\begin{equation}
   \frac{\rho_2}{\rho_1} \sim \min\left\{\frac{\Gamma}{\delta \Gamma},\frac{t}{t_0}\right\} \, .
\end{equation}

\begin{figure}
\begin{center}
\begin{tabular}{c}
\resizebox{0.9\hsize}{!}{\includegraphics{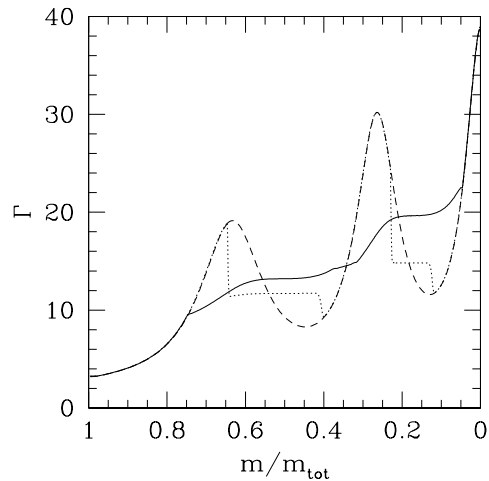}}
\end{tabular}
\end{center}
\caption{Hydrodynamic simulation of a spherical 
outflow undergoing internal shocks during its expansion.
The flow is injected cold with a constant kinetic power during $t_w = 10$~s at a radius 
$R_0 = 0.1 \, c t_w $.
The figure shows the Lorentz factor profile as a function of normalized Lagrangian mass coordinate,
at three moments of time: the initial (injected) state (dashed line),
during the internal shock phase at $t_{\rm lab} =800$~s (dotted line),
and after internal shocks ended at $t_{\rm lab} =10^5$~s (solid line).
At $t_{\rm lab} = 10^5$ s, well after the internal shock phase, the outflow has cooled adiabatically and entered a ballistic phase.
As a result of internal shocks, two large portions of the ejecta have a uniform Lorentz factor with $\delta \Gamma / \Gamma \la 0.01$.
We have checked that adding a fast variability component in the outflow (leading to several generations of internal shocks) 
does not significantly change the final Lorentz factor distribution.
}
\label{fig_hydro}
\end{figure}

An important feature of the post-internal-shock flows is the presence of a few 
``plateaus'' in the profile of $\Gamma(r)$ or $\Gamma(m)$ where $m$ is the Lagrangian 
mass coordinate in the flow. These plateaus contain a large mass and eventually form 
dense massive shells because of their small $\delta\Gamma$. We have observed and
studied this effect in outflows with various initial $\Gamma(m)$ using detailed 
hydrodynamical simulations. 
We found that that the plateaus typically have a small $\delta \Gamma / \Gamma$ 
of a few per cent, even in the case of adiabatic shocks (an example is shown in Figure~\ref{fig_hydro}).
The simulations have been performed using a one-dimensional (spherically symmetric)
hydrodynamic code with  a second-order HLLC approximate Riemann solver \citep{mignone_2005}
on a moving mesh \citep{duffell_2011}.

Similar (but exactly flat) plateaus are produced by simplified simulations in which the 
outflow is represented by a large number of discrete shells that interact by direct collisions 
\citep{daigne_1998}. In this model, the ejecta is assumed to be cold and all pressure waves
are neglected. This simplification is not crucial as it gives results similar to the detailed hydrodynamical simulations (cf. Daigne \& Mochkovitch 2000). 
Below we use the simplified model and extend the simulation to include the external
blast wave and the reverse shock propagation through the outflow.


\section{Flares: a simplified model}
\label{sect_dissPow}

\begin{figure*}
\begin{center}
\begin{tabular}{cc}
\resizebox{0.47\hsize}{!}{\includegraphics{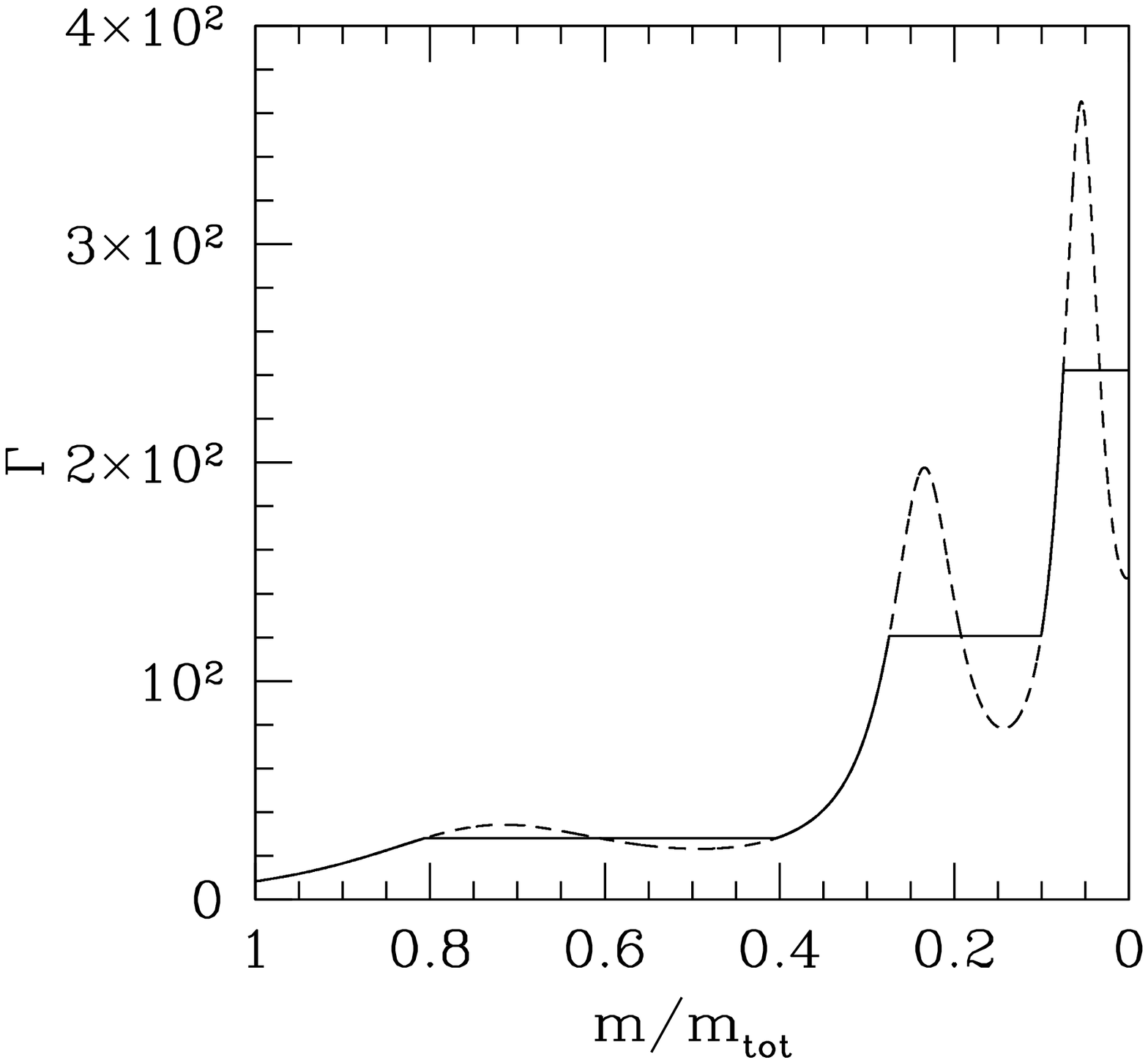}}
\resizebox{0.47\hsize}{!}{\includegraphics{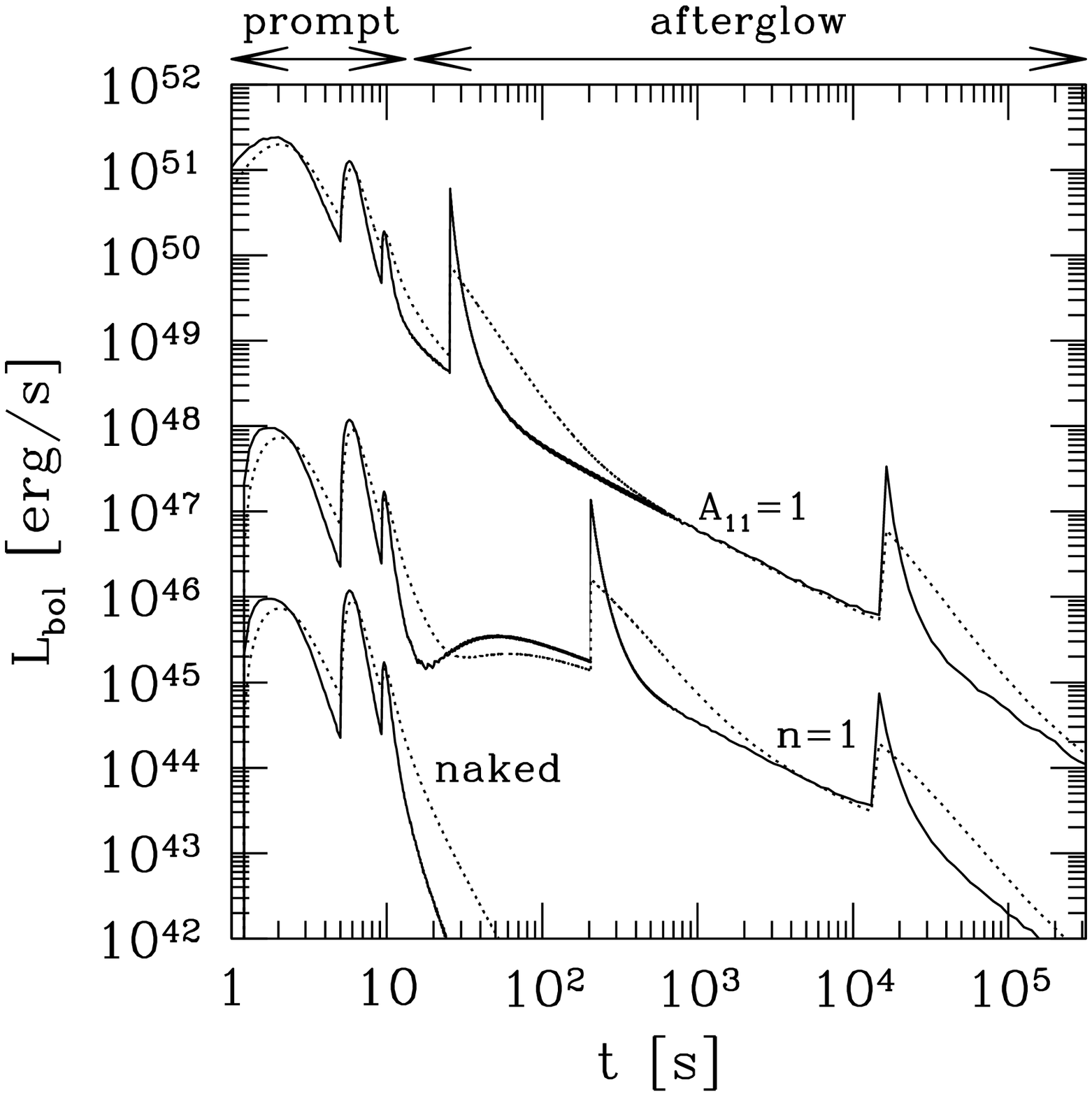}}\\
\end{tabular}
\end{center}
\caption{Left: Lorentz factor distribution as a function of Lagrangian mass coordinate; the front edge of the outflow is at $m=0$. 
The initial distribution $\Gamma(m)$ is shown by the dashed curve, and the final distribution (after internal shocks) is shown by the solid curve.
Right: bolometric light-curves (assuming a radiative efficiency $\erad = 0.1$) calculated for different ambient media (dotted lines). 
The wind density $\rho$ is parametrized by $A_{11}$ as follows: $\rho R^2=  10^{11} A_{11} \ \mathrm{g\ cm^{-1}}$.
In the uniform medium case, $n$ is the density in $\mathrm{cm^{-3}}$.
From top to bottom: 
wind medium with $A_{11}=1$, 
uniform medium with $n=1$ ($\lbol$ has been divided by $10^{3}$),
and the case of a ``naked'' burst (no ambient medium; $\lbol$ has been divided by $10^{5}$).
The solid lines show the same light-curves for the case of an anisotropic emission,
which is beamed within an angle $\theta_{\rm beam} \approx 60^\mathrm{\circ}$ in the plasma frame.
In the case of the naked burst, the three early pulses (corresponding to the three pairs of internal shocks that initially propagate within the ejecta)
are followed by high-latitude (off-axis) emission.
} 
\label{fig_lfDist}
\end{figure*}

We used the simplified hydrodynamical simulations to follow the reverse shock (RS)
propagation through the outflow and its interaction with the dense shells.
A sample model is shown in Figure~\ref{fig_lfDist}.
In this example, the explosion is assumed to produce an outflow of variable Lorentz factor distribution and duration $t_{\rm w} = 10$~s. 
The injected kinetic power ${\dot E}_{\rm K}=10^{53}$ erg s$^{-1}$  is assumed
to be constant.
Figure~2 shows how the initial distribution of $\Gamma$ 
is changed after the internal shocks.
Most of the outflow mass concentrates into three shells with Lorentz factors $\Gamma \approx 240$, $120$ and $30$ 
carrying about 10, 20, and 40 per cent of the total mass. 
The last 30 per cent of the outflow have not been affected by internal shocks.

We followed the outflow to much larger radii where the RS
eventually crosses all three shells. The dynamics of the reverse shock depends on the
external density.
We have calculated two cases: a uniform external medium $\rho=const$ and a wind-like 
medium $\rho=A r^{-2}$, where $A$ is a constant which depends on the mass-loss rate of 
the GRB progenitor. 
Its typical range for galactic Wolf-Rayet stars is $10^{11}-10^{12} \ \mathrm{g \ cm^{-1}}$ \citep{crowther_2007}.
Recent modelling of optical and GeV flashes in GRBs suggest that $A \sim 10^{11} \ \mathrm{g \ cm^{-1}}$ is typical for GRB progenitors 
(\citealt{hascoet_2014b}; Hasco\"et, Vurm \& Beloborodov, to be submitted).

\subsection{Bolometric light curve}

The simulation gives the power dissipated in the RS and
its bolometric luminosity with two assumptions:
\textit{(i)} that a fraction $\ee\sim 0.1$ of the dissipated energy is injected in 
shock-accelerated electrons, and \textit{(ii)} that these electrons promptly radiate 
their energy (``fast-cooling'' regime).
To find the bolometric light curve received by a distant observer 
we transform the emission from the flow frame to the observer frame and take into account
the spherical curvature of the emitting shells. The radiation received by the 
observer from the RS is shown in Figure~2.
In this figure, we also included the prompt radiation produced by the internal shocks,
assuming the same $\ee=0.1$.

When the reverse shock goes through the dense shells 
there is a sudden rise of dissipated power producing a flare in the light curve.
The immediate rise of luminosity to the peak of the flare in Figure~2 is an artefact of 
the simplified model, which does not resolve the propagation of the reverse shock through the 
dense shell. An accurate hydrodynamical model would give a slower but still a steep rise; 
it is discussed and estimated in Section~\ref{subsect_timescales} below.
 
In our example with three dense shells, two flares are produced;
the crossing of the first shell (at $\Gamma \approx 240$) corresponds 
to the initial rise of dissipated power from the reverse shock.
Using the approximate expression for the flare radius 
$R_{\rm f} \sim \Gamma^2 c \tf$, one can estimate
the observed time of the flare,
\begin{eqnarray}
\label{eq_ta}
t_{\rm f}
    =\left({3\,E\over 4\pi\,n\,m_p\,c^5\,\Gamma^8}\right)^{1/3}
       \approx 840 \,\left({E_{54}\over n}\right)^{1/3}\,\Gamma_2^{-8/3}\ {\rm s} 
  \quad &{\rm (uniform),}  & \\
   t_{\rm f}={E\over 4 \pi\,A\,c^3\,\Gamma^4}
       \approx 
       300
       \,\left({E_{54}\over 
       A_{11}
       }\right)\,\Gamma_2^{-4}\ {\rm s} 
   \quad &{\rm (wind).}&
\end{eqnarray}
Here $\Gamma$ is the Lorentz factor of the massive shell
($\Gamma_2=\Gamma/100$), $E$ is the kinetic energy of the ejecta 
that has already crossed the reverse shock ($E_{54}=E/10^{54}$ erg), $n$ 
is the density of the uniform medium in cm$^{-3}$ and 
$A_{11}$ is the density parameter of the wind medium in units of $10^{11} \ \mathrm{g \ cm^{-1}}$.
One can see that the flare can be significantly delayed and this delay is sensitive to 
the Lorentz factor of the massive shell.

We also observe that the flares produced by this mechanism share two key 
features with the observed X-ray flares: 
{\it (i)} after the flare the light curve returns to the pre-flare decaying afterglow and 
{\it (ii)} the flare duration $\Delta t_{\rm f}$ is proportional to its time of occurrence $t_{\rm f}$.

The simulation shown in Figure~2 assumes fast cooling at all times, which gives 
a constant radiative efficiency equal to $\ee$. 
The flux enhancement during the flare can be even higher if before and after the flare the 
emission is in the slow-cooling regime. Then the radiative efficiency can be significantly
increased during the flare.

\subsection{Effect of anisotropic emission}
\label{subsect_anisotropy}

In the simplest model, which assumes isotropic emission in the fluid frame, 
the characteristic temporal width of the flare is $\dtf\sim \tau$ where
\begin{equation}
\tau= \frac{R}{2\Gamma^2 c^2} \sim \tf.
\end{equation}
Then the ratio $\dtf/\tf$ is close to unity, greater 
than the typical observed ratio $\dtf/\tf\sim 0.1-0.3$. 

Significant anisotropy is, however, expected and 
supported by the fast luminosity variations observed in GRB afterglows \citep{beloborodov_2011}.
The effect of anisotropy on the flare light curve is demonstrated by the following 
simple model.

For the moment, let us picture the encounter of
the reverse shock with the dense shell as an instantaneous flash of energy $E_{\rm f}$
at a radius $R$. The bolometric light curve received from such a flash 
is given by \citep{beloborodov_2011},
\begin{equation}
 L(t)={2 E_{\rm f}A(\theta)\over \tau}\left(1+{t-\tf \over \tau}\right)^{-3}
  \quad \rm (t>\tf), 
\end{equation}
where
\begin{equation}
   \cos\theta=\frac{\tau-(t-\tf)}{\tau+(t-\tf)},
\end{equation}
$\theta$ is the angle with respect to the radial direction in the fluid frame, 
and $A(\theta)$ describes the angular distribution of emission; for isotropic 
emission $A(\theta)=1$. The magnetic field in GRB shocks is in the shock plane,
and their synchrotron radiation is anisotropic even when the emitting electrons
are isotropic; in this case $A(\theta)=(3/4)(1+\cos^2\theta)$. Anisotropy of the electron
distribution can further enhance the anisotropy of the emitted radiation, which significantly 
shortens the duration of the flare (see Figure~3 in \citealt{beloborodov_2011}), 
leading to a characteristic duration $\dtf \ll\ \tf$.
The shorter duration corresponds to a steeper decay of luminosity after the peak,
with initial temporal index $\alpha\sim 3 \tf/\dtf$.

As an illustration, in \fig{fig_lfDist} we show 
the model where the emission is moderately limb-darkened in the plasma frame,
so that it is beamed within an angle $\theta_{\rm beam} \approx 60^\mathrm{\circ}$ 
(see Equation~9 in \citealt{beloborodov_2011}). Then $\dtf\sim 0.3 \tf$ and the 
resulting light curves resemble the observed X-ray flares.


\section{Shock reflection from the dense shell}
\label{subsect_timescales}

\begin{figure}
\begin{center}
\begin{tabular}{c}
\resizebox{1.\hsize}{!}{\includegraphics{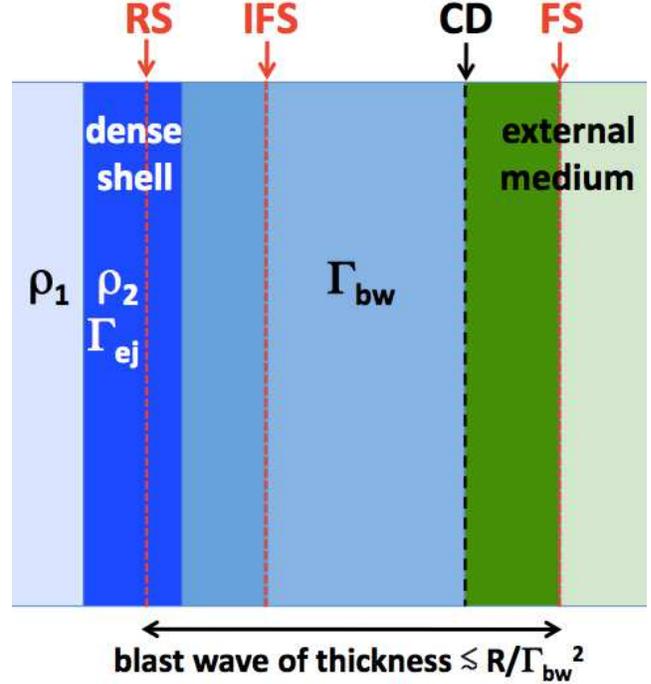}}
\end{tabular}
\end{center}
\caption{Schematic illustration of the 
reverse shock (RS) encounter with a thin massive shell.
The RS becomes slow as it enters the dense material and a new 'internal forward shock' (IFS) 
is created, which propagates away from the dense shell into the blast wave material 
previously shocked by the RS.
The figure also shows the forward shock (FS) in the external medium and 
the contact discontinuity (CD) that separates the ejecta from the external medium inside the blast wave.
} 
\label{fig_cartoon}
\end{figure}

In the simplified model of Section~3, the RS crossing of the thin dense shell 
instantaneously dissipates 
part of 
the shell kinetic energy, and the observed light curve 
of the flash is determined by the curvature effect. In this section, we describe a more 
realistic picture of the RS interaction with the dense shell. 
We also estimate the timescales for the rise and duration of the dissipation event.

Two main effects should be taken into account:
(i) The RS is greatly slowed down as it enters the dense gas,
and it takes a finite time (estimated below)  for the RS to cross the shell.
(ii) The RS by itself does not efficiently decelerate the shell. 
The shell continues to plow through the ejecta that was previously accumulated in the 
blast wave. This ploughing launches a new internal shock inside the blast wave  
(Figure~\ref{fig_cartoon}). Effectively, the RS splits in two when it enters the dense shell. 
One (slow) shock continues to propagate into the dense 
shell and the other (fast) shock is reflected and launched back into the blast wave material.
The fast shock dissipates most of the kinetic energy of the massive shell
measured in the rest-frame of the blast wave.
Below we call it `internal forward shock' (IFS). 
It is the IFS that emits the X-ray flare.

The velocity of RS propagation through the dense shell, 
$\betars=v_{\rm RS}/c\ll 1$, may be estimated using the approximate pressure balance
between the RS and IFS. 
Let $\rho_2$ be the proper density of the shell just before it is reached by the RS.
Assuming that the shell is cold and weakly magnetized, its pressure behind the RS jumps to
\begin{equation}
\label{eq_p2}
\pRS = \frac{3}{4} \,\betars^2\,\rho_2 c^2.
\end{equation} 
Now, let us estimate the pressure behind the IFS, $\pIFS$. 
The dense shell acts like a wall and the Lorentz factor downstream of the IFS 
is almost equal to the shell Lorentz factor $\Gej$. The relative Lorentz factor
of the upstream and downstream is
\begin{equation}
\label{eq_gr}
\Gr = \frac{1}{2}\left( \frac{\Gamma_{\rm ej}}{\Gamma_{\rm bw}} + \frac{\Gamma_{\rm bw}}{\Gamma_{\rm ej}} \right),
\end{equation}
where $\Gbw$ is the Lorentz factor of the blast-wave material  upstream of the IFS,
and we have used $\Gamma_{\rm ej}\gg 1$ and $\Gamma_{\rm bw}\gg 1$.
Gas pressure may be written as $p\approx (\rho u/3)(1+c^2/u)$
\citep{beloborodov_2006},
where $\rho$ is the density and $u$ is the energy 
per unit rest mass.
The gas in the downstream of IFS has been shocked twice:
first by the RS and then by the IFS, both times with the same relative Lorentz factor
$\Gr$. Therefore, $u\approx \Gr^2 c^2$ and $\rho\approx(4\Gr)^2\rho_1$, where $\rho_1$ is the outflow 
proper density {\em prior to} the crossing by the RS and IFS. In the 
estimate for $\rho$ we used the compression factor of $4\Gr$ for a strong adiabatic shock.
This gives
\begin{equation}
\label{eq_p1}
  \pIFS\approx \frac{16}{3} \left( \Gr^{4} -1\right) \rho_1 c^2.
\end{equation}
The approximate pressure balance $\pRS\sim \pIFS$ gives an estimate for $\betars$,
\begin{equation}
  \betars \approx
  \frac{8}{3} \left[(\Gr^4 -1)\frac{\rho_1}{\rho_2}\right]^{1/2}. 
\end{equation}

Using the obtained $\betars$, we estimate the time it takes the RS to cross
the dense shell, $t_{\rm cross}$. The shell thickness is comparable to
\begin{equation}
    \Delta_{\rm sh} =\eta \left( \frac{\rho_1}{\rho_2} \right) \Delta_{\rm bw} \, ,
\end{equation}
where $\Delta _{\rm bw} \la R/\Gamma_{\rm bw}^2 \approx ct$ is the blast wave thickness, 
and $\eta = m_{\rm sh} / m_{\rm bw} \la 1$ is the ratio between the mass of the incoming dense shell,  $m_{\rm sh}$,
and the outflow mass already accumulated in the blast wave,  $m_{\rm bw}$.
Then we find
\begin{equation}
\label{eq_tcross}
   t_{\rm cross} \approx \frac{\Delta_{\rm sh}}{2c \beta_{\rm RS}} 
\la 0.2\, \eta\, \left( \frac{\rho_1}{\rho_2} \right)^{1/2} 
\left( \Gr^4 -1\right)^{-1/2} t.
\end{equation}
Typical expected $\rho_2/\rho_1>30$ implies a short $\tcross$.

The shell Lorentz factor is reduced from $\Gamma_{\rm ej}$ to $\Gamma_{\rm bw}$ when it sweeps up the rest-mass
\begin{equation}
   m_\star\sim\frac{\msh}{\Gr^2} \, ,
\end{equation}
where we took into account that the gas ahead of the shell has been heated by the RS,
and so its inertial mass is increased by the factor of $\Gamma_{\rm rel}$.
Assuming a relativistic IFS speed $\sim c$, this gives the following estimate for the observed 
duration of the shell deceleration:
\begin{equation}
\label{eq_tdiss}
t_{\rm diss} \sim 
     \frac{m_*}{m_{\rm bw}} \frac{\Delta_{\rm bw}}{c} 
         \left( \frac{\Gamma_{\rm bw}}{\Gamma_{\rm ej}} \right)^2 \, 
      \la \frac{m_*}{m_{\rm bw}}  
       \left( \frac{\Gamma_{\rm bw}}{\Gamma_{\rm ej}} \right)^2 t \, .
\end{equation}

The IFS may be mildly relativistic, $\Gr\sim 2$, if the flares occur 
long after the formation of the ballistic outflow ($t\gg t_0$). 
If the blast wave dynamics between the flares 
is approximated  as self-similar, one finds  
$\Gej/\Gbw = \sqrt{2}$ and $2$ in the case of the wind and uniform external medium respectively.

The time $\tcross$ is associated with the rise of the
energy injection event and $\tdiss$ 
determines its duration. The above results give typical $\tcross\ll\tdiss\ll t$. For short 
$\tdiss$ the observed duration of the flare will be determined by the curvature (and 
anisotropy) effects, as described in Section~3.2.  Here we note an additional effect which 
reduces the observed duration of the flare. While the pre-flare afterglow is emitted by gas 
moving with $\Gbw$, the flare is emitted by gas moving with $\Gej>\Gbw$ (the gas behind 
the IFS). Therefore, even without the anisotropy effect, the curvature timescale for the flare 
is compressed by the factor of $\Delta \tf/\tf\sim (\Gbw/\Gej)^2$.

\section{Conclusions}
\label{sect_discussion}

This Letter suggests that GRB outflows naturally develop a few dense shells with relatively low Lorentz factors, 
which leads to a new explanation for the X-ray flares. 
The dense shells give sudden energy injections to the GRB afterglow. 
We examined the dynamics of the injection event and showed that it generates a new internal shock behind the blast wave, 
which produces a short-lived flare. 
This mechanism provides a natural explanation for the fast rise of the X-ray flares 
and their observed duration $\Delta t_f / t_f \sim 0.1-0.3$. 

The proposed scenario assumes two radiative properties of the shock: 
(i) a fraction of the shock energy is transferred to non-thermal accelerated electrons capable of emitting synchrotron X-rays, 
and (ii) the electrons radiate X-rays in the fast cooling regime. 
Detailed modeling of hydrodynamics and radiative properties of the flares are deferred to a future work. 
As described in Section 4, the flare is emitted by gas that has been shocked twice and has high density and pressure. 
This gas can also possess strong magnetic fields and can radiate more efficiently compared with the pre-flare afterglow. 
This will help to satisfy the conditions (i) and (ii). 

\section*{Acknowledgments}

It is a pleasure to thank H. van Eerten and A. I. MacFadyen for useful discussions on hydrodynamics simulations.
This work has been partially supported by the Programme National Hautes
Energie (PNHE) and the French Space Agency (CNES) .
AMB was supported by NSF grant AST-1412485 and NASA Swift Cycle 10 grant NNX14AI94G.

\bibliographystyle{apj}  
\bibliography{main}
\label{lastpage}

\end{document}